\documentclass[12pt]{article}

\usepackage{graphicx}
\usepackage{scicite}
\usepackage{times}
\usepackage{natbib}
\usepackage{fancyhdr}

\renewcommand{\headrulewidth}{0.5pt}
\renewcommand{\footrulewidth}{0pt}

\newenvironment{sciabstract}{%
\begin{quote} \bf}
{\end{quote}}

\title{Functional cartography\\of complex metabolic networks}
\author{Roger Guimer\`a and Lu\'{\i}s A. Nunes Amaral\\
\normalsize{NICO and Dept. Chemical and Biological Engineering} \\
\normalsize{Northwestern University, Evanston, IL 60208, USA}\\ \\
}

\date{}

\renewcommand{\baselinestretch}{1.5}

\begin{document}

\maketitle
%


\begin{sciabstract}
High-throughput techniques are leading to an explosive growth in the
size of biological databases and creating the opportunity to
revolutionize our understanding of life and disease. Interpretation of
these data remains, however, a major scientific challenge. Here, we
propose a methodology that enables us to extract and display
information contained in complex
networks~\cite{amaral00,albert02,amaral04}. Specifically, we
demonstrate that one can (i) find functional
modules\cite{hartwell99,girvan02} in complex networks, and (ii)
classify nodes into universal roles according to their pattern of
intra- and inter-module connections. The method thus yields a
``cartographic representation'' of complex networks. Metabolic
networks \cite{jeong00,wagner01,ma03} are among the most challenging
biological networks and, arguably, the ones with more potential for
immediate applicability\cite{hatzimanikatis04}. We use our method to
analyze the metabolic networks of twelve organisms from three
different super-kingdoms. We find that, typically, 80\% of the nodes
are only connected to other nodes within their respective modules, and
that nodes with different roles are affected by different evolutionary
constraints and pressures. Remarkably, we find that low-degree
metabolites that connect different modules are more conserved than
hubs whose links are mostly within a single module.
\end{sciabstract}


If one is to extract the significant information from the topology of
a large complex network, the knowledge of the role of each node is of
crucial importance. A cartographic analogy is helpful to illustrate
this point. Consider the network formed by all cities and towns in a
country---the nodes---and all the roads that connect them---the
links. It is clear that a map in which each city and town is
represented by a circle of fixed size and each road is represented by
a line of fixed width is hardly useful. Rather, real maps emphasize
capitals and important communication lines so that one can obtain
scale-specific information at a glance. Similarly, it is difficult, if
not impossible, to obtain information from a network with hundreds or
thousands of nodes and links, unless the information about nodes and
links is conveniently summarized. This is particularly true for
biological networks.

Here, we propose a methodology, which is based on the connectivity of
the nodes, that yields a ``cartographic representation'' of a complex
network.  The first step in our method is to identify the functional
modules\cite{hartwell99,girvan02} in the network. In the cartographic
picture, modules are analogous to countries or regions, and enable a
coarse-grained, and thus simplified, description of the network. Then,
we classify the nodes in the network into a small number of {\it
system-independent\/} ``universal roles.''

\bigskip
\noindent
{\it Modules.}
It is a matter of common experience that social networks have
communities of highly interconnected nodes that are less connected to
nodes in other communities. Such modular structures have been reported
not only in social
networks~\cite{girvan02,guimera03,newman03,arenas04}, but also in food
webs~\cite{krause03} and biochemical
networks~\cite{hartwell99,ravasz02,holme03,papin04}. It is widely
believed that the modular structure of complex networks plays a
critical role in their
functionality~\cite{hartwell99,ravasz02,papin04}. There is therefore a
clear need to develop algorithms to identify modules
accurately~\cite{girvan02,newman03,eriksen03,newman04,radicchi04,donetti04}.

\begin{figure} 
\centerline{
\includegraphics*[height=.7\textwidth]{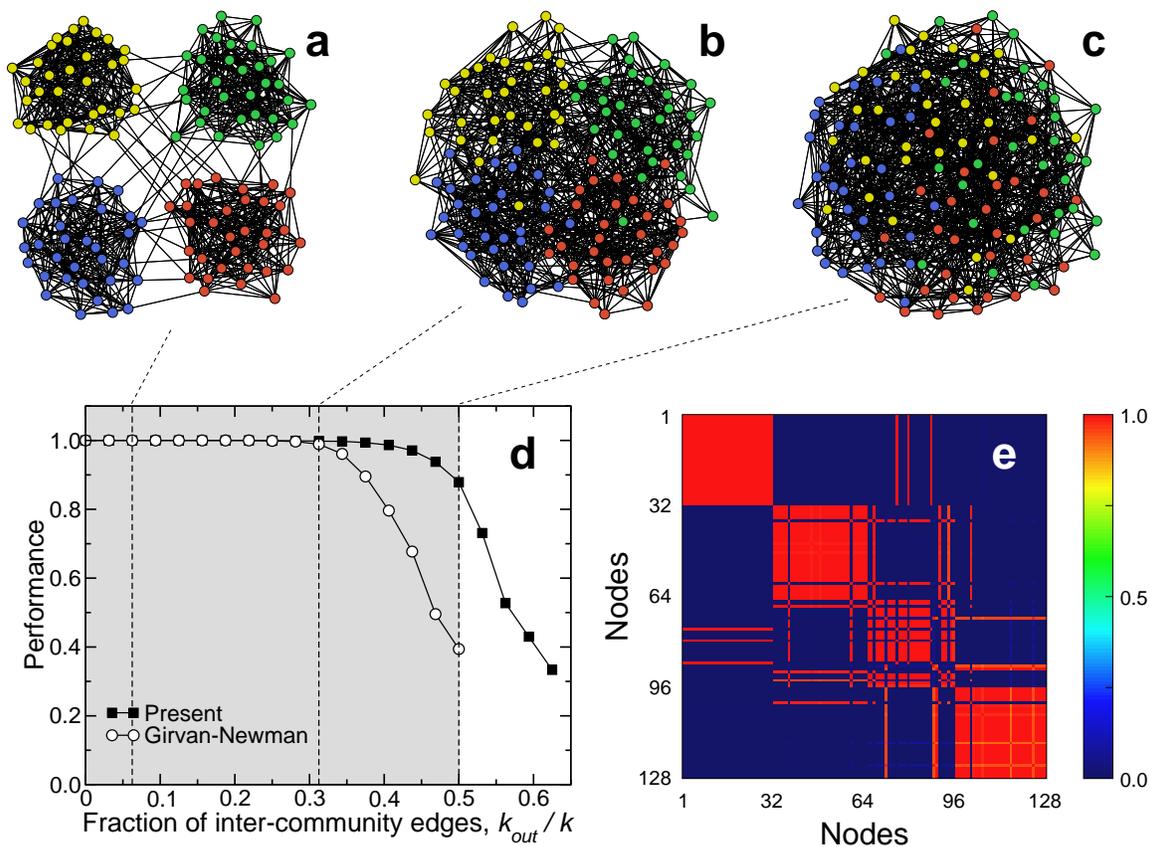}
}
\renewcommand{\baselinestretch}{1.0}
\caption{
Performance of module identification methods.
To test the performance of the method, we build ``random networks''
with known module structure. Each test network comprises 128 nodes
divided into 4 modules of 32 nodes. Each node is connected to the
other nodes in its module with probability $p_{i}$, and to nodes in
other modules with probability $p_{o}<p_{i}$. On average, thus, each
node is connected to $k_{out}=96\,p_{o}$ nodes in other modules and to
$k_{in}=31\,p_{i}$ in the same module. Additionally, $p_{i}$ and
$p_{o}$ are selected so that the average degree of the nodes is
$k=16$. We display networks with: {\bf a,} $k_{in}=15$ and
$k_{out}=1$; {\bf b,} $k_{in}=11$ and $k_{out}=5$; and {\bf c,}
$k_{in}=k_{out}=8$.
{\bf d,} The performance of a module identification algorithm is
typically defined as the fraction of correctly classified nodes. We
compare our algorithm to the Girvan-Newman
algorithm~\cite{girvan02,newman04}, which is the reference algorithm
for module identification ~\cite{newman03,newman04,radicchi04}. Note
that our method is 90\% accurate even when half of a node's links are
to nodes in outside modules.
{\bf e,} Our module-identification algorithm is stochastic, so
different runs yield, in principle, different partitions. To test the
robustness of the algorithm, we obtain 100 partitions of the network
depicted in {\bf c} and plot, for each pair of nodes in the network,
the fraction of times that they are classified in the same module. As
shown in the figure, most pairs of nodes are either always classified
in the same module (red) or never classified in the same module (dark
blue), which indicates that the solution is robust. 
}
\label{f-perf-mod}
\end{figure}

We identify modules by maximizing the network's {\it
modularity}~\cite{newman03,newman04,guimera04c} using simulated
annealing~\cite{kirkpatrick83} (see Methods). Simulated annealing
enables us to carry out an exhaustive search and to minimize the
problem of finding sub-optimal partitions. It is noteworthy that, in
our method, one does not need to specify a priori the number of
modules; rather, this number is an outcome of the algorithm. Our
algorithm, which significantly outperforms the best algorithm in the
literature, is able to reliably identify modules in a network whose
nodes have as many as 50\% of their connections outside their own
module (Fig.~\ref{f-perf-mod}).

\bigskip
\noindent
{\it Roles in modular networks.}
It is plausible to surmise that the nodes in a network are connected
according to the {\it role\/} they fulfill. This fact has been long
recognized in the analysis of social
networks~\cite{wasserman94}. For example, in a classical
hierarchical organization, the CEO is not directly connected to plant
employees but is connected to the members of the board of
directors. Importantly, such a statement holds for virtually any
organization, that is, the role of CEO is defined irrespective of the
particular organization one considers.

We propose a new method to determine the role of a node in a complex
network. Our approach is based on the idea that nodes with the same
role should have similar topological properties\cite{guimera??e} (see
Supplementary Information for a discussion on how our approach relates
to previous work). We hypothesize that the role of a node can be
determined, to a great extent, by its {\it within-module degree} and
its {\it participation coefficient}, which define how the node is
positioned in its own module and with respect to other
modules\cite{rives03,han04} (see Methods). These two properties are
easily computed once the modules of a network are known.

The within-module degree $z_i$ measures how ``well-connected'' node
$i$ is to other nodes in the module. High values of $z_i$ indicate
high within-module degrees and vice versa. The participation
coefficient $P_i$ measures how ``well-distributed'' the links of node
$i$ are among different modules. The participation coefficient $P_i$
is close to one if its links are uniformly distributed among all the
modules and zero if all its links are within its own module.

\begin{figure} 
\centerline{
\includegraphics*[width=\textwidth]{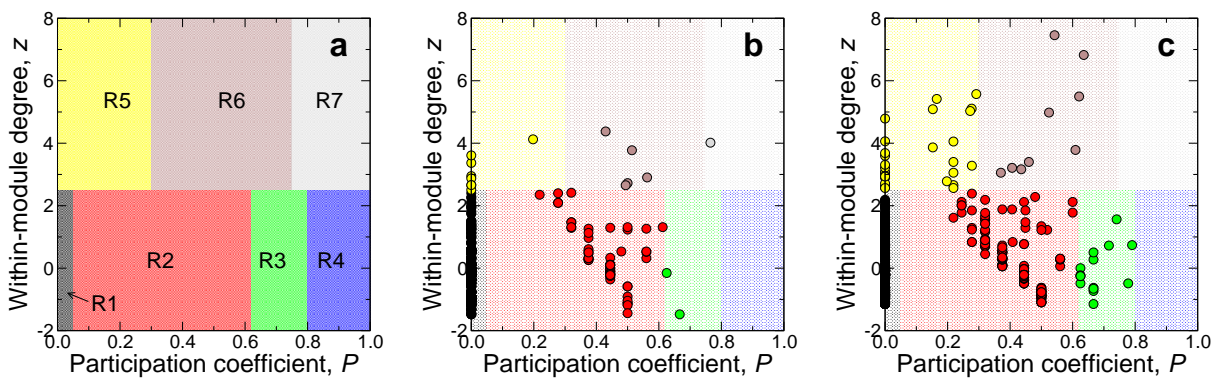}
}
\renewcommand{\baselinestretch}{1.0}
\caption{
Roles and regions in the $zP$ parameters-space. {\bf a,} Each node in
a network can be characterized by its within-module degree and its
participation coefficient (see Methods for definitions.) We classify
nodes with $z\ge 2.5$ as module hubs and nodes $z<2.5$ as non-hubs. We
find that non-hub nodes can be naturally assigned into four different
roles: (R1) {\it ultra-peripheral nodes}, i.e., nodes with all its
links within their module; (R2) {\it peripheral nodes}, i.e., nodes
with most links within their module; (R3) {\it non-hub connector
nodes}, i.e., nodes with many links to other modules; and (R4) {\it
non-hub kinless nodes}, i.e., nodes with links homogeneously
distributed among all modules. We find that hub nodes can be naturally
assigned into three different roles: (R5) {\it provincial hubs}. i.e.,
hub nodes with the vast majority of links within their module; (R6)
{\it connector hubs}, i.e., hubs with many links to most of the other
modules; and (R7) {\it kinless hubs}, i.e., hubs with links
homogeneously distributed among all modules. (Supplementary
Information.)
{\bf b,} Metabolite role determination for the metabolic network {\it
E. coli}, as obtained from the MZ database. Each metabolite is
represented as a point in the $zP$ parameters-space, and is colored
according to its role.
{\bf c,} Same as {\bf b} but for the complete KEGG database.
}
\label{f-roledef}
\end{figure}

We define heuristically seven different ``universal roles,'' each
defined by a different region in the $zP$ parameters-space
(Fig.~\ref{f-roledef}). According to the within-module degree, we
classify nodes with $z \ge 2.5$ as module hubs and nodes $z<2.5$ as
non-hubs.  Both hub and non-hub nodes are then more finely
characterized by using the values of the participation coefficient
(see Supplementary Information for a detailed justification of this
classification scheme, and for a discussion on possible alternatives).

We find that non-hub nodes can be naturally divided into four
different roles: (R1) {\it ultra-peripheral nodes}, i.e., nodes with
all its links within their module ($P \le 0.05$); (R2) {\it peripheral
nodes}, i.e., nodes with most links within their module ($0.05<P \le
0.62$); (R3) {\it non-hub connector nodes}, i.e., nodes with many
links to other modules ($0.62<P \le 0.80$); and (R4) {\it non-hub
kinless nodes}, i.e., nodes with links homogeneously distributed among
all modules ($P>0.80$). We find that hub nodes can be naturally
divided into three different roles: (R5) {\it provincial hubs}. i.e.,
hub nodes with the vast majority of links within their module ($P \le
0.30$); (R6) {\it connector hubs}, i.e., hubs with many links to most
of the other modules ($0.30<P \le 0.75$); and (R7) {\it kinless hubs},
i.e., hubs with links homogeneously distributed among all modules
($P>0.75$).

\begin{figure}
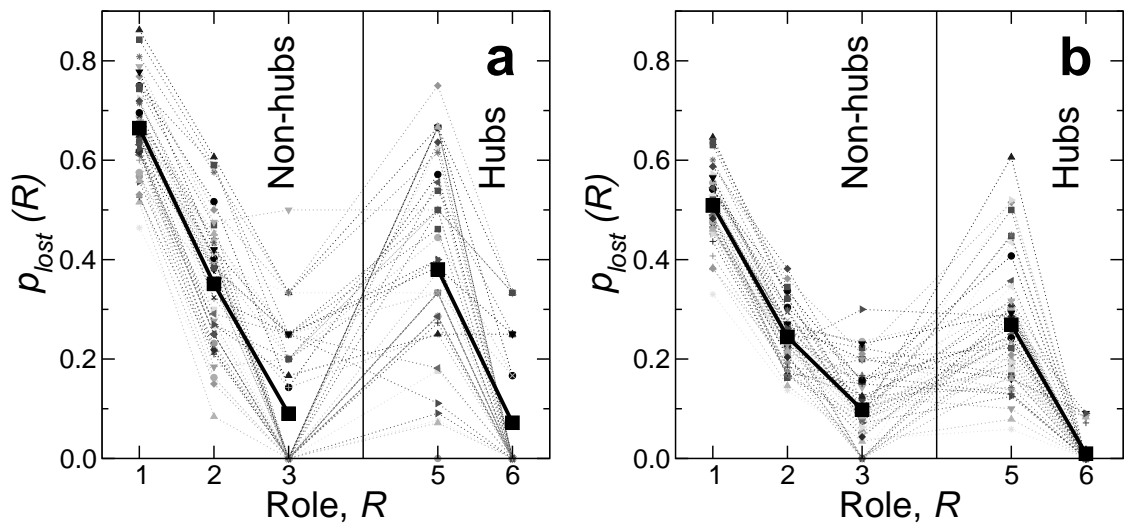

\centerline{
\includegraphics*[width=0.45\textwidth]{roleconserv-Ma}\quad
\includegraphics*[width=0.45\textwidth]{roleconserv}
}
\renewcommand{\baselinestretch}{1.0}
\caption{
Roles of metabolites and inter-species conservation. To quantify the
relation between roles and conservation, we calculate the loss rate
$p_{\rm lost}(R)$ of each metabolite (see Methods).
Each thin line in the graph corresponds to a comparison between two
species. Since we are interested in metabolites that are present in
some species but missing in others, metabolic networks of species
within the same super-kingdom---bacteria, eukaryotes, and
archaea---are usually too similar to provide statistically sound
information, especially for roles containing only a few
metabolites. Therefore, we consider in our analysis only pairs of
species that belong to different super-kingdoms. The thick line is the
average over all pairs of species.
The loss rate $p_{\rm lost}(R)$ is maximum for ultra-peripheral (R1)
nodes and minimum for connector hubs (R6). Remarkably, provincial hubs
(R5) have a significantly and consistently higher $p_{\rm lost}(R)$
than non-hub connectors (R3), even though the within-module degree and
the total degree of provincial hubs is larger.
Note that, out of the total 48 pair comparisons, only in two cases
$p_{\rm lost}(R)$ is lower for provincial hubs than for non-hub
connectors, while the opposite is true in 44 cases.
{\bf a,} Results obtained for the MZ database and {\bf b,} the
complete KEGG database.
}
\label{f-conservation}
\end{figure}

\bigskip
\noindent
{\it Cartographic representation of metabolic networks.}
To test the applicability of our approach to complex biological
networks, we consider the metabolic
network~\cite{jeong00,wagner01,ravasz02,ma03,hatzimanikatis04} of
twelve organisms: four bacteria ({\it E. coli}, {\it B. subtilis},
{\it L. lactis}, and {\it T. elongatus}), four eukaryotes ({\it
S. cerevisiae}, {\it C. elegans}, {\it P. falciparum}, and {\it
H. sapiens}), and four archaea ({\it P. furiosus}, {\it A. pernix},
{\it A. fulgidus}, and {\it S. solfataricus}). In metabolic networks,
nodes represent metabolites and two nodes $i$ and $j$ are connected by
a link if there is a chemical reaction in which $i$ is a substrate and
$j$ a product, or vice versa. In our analysis, we use the database
developed by Ma and Zeng~\cite{ma03} (MZ) from the Kyoto Encyclopedia
of Genes and Genomes~\cite{kanehisa00} (KEGG). Importantly, the
results we report are not altered if we consider the complete KEGG
database instead (Figs.~\ref{f-roledef}c and \ref{f-conservation}b,
and Supplementary Information).

\begin{figure}
\centerline{\includegraphics*[width=\textwidth]{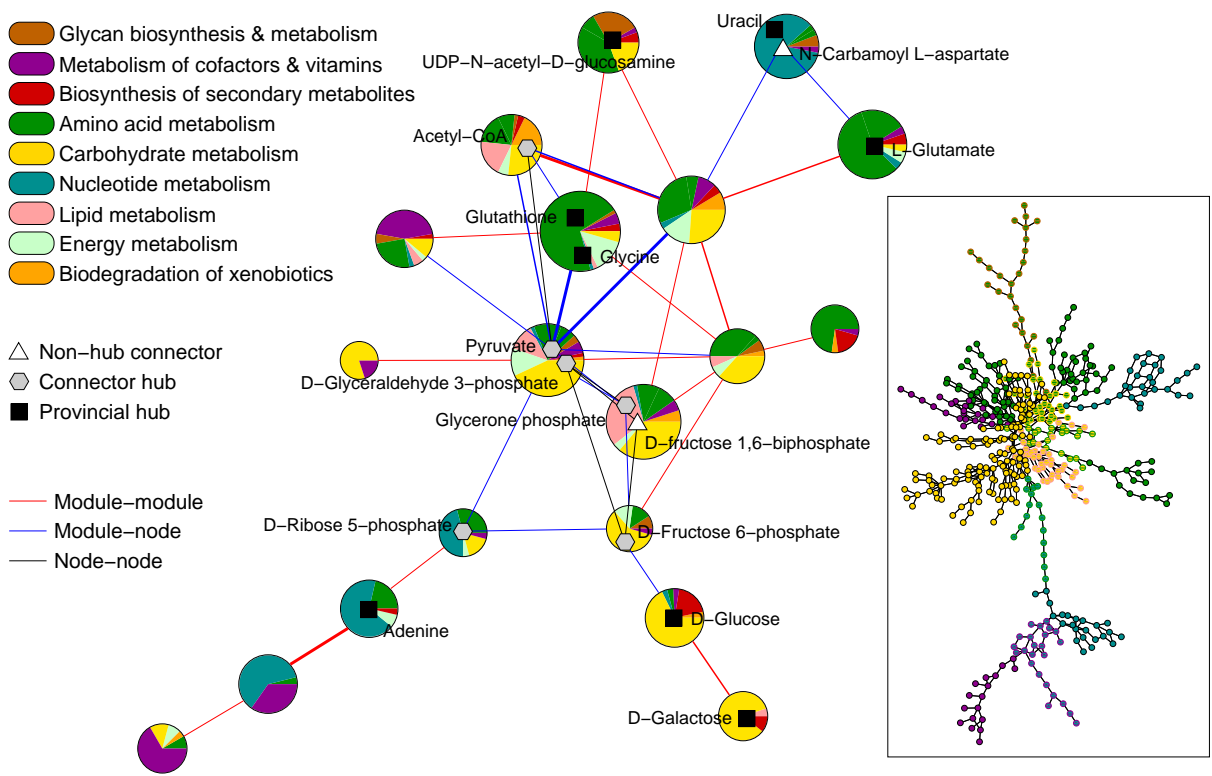}}
\renewcommand{\baselinestretch}{1.0}
\caption{
``Cartographic representation'' of the metabolic network of {\it
E. coli}.
Each circle represents a module and is colored according to the KEGG
pathway classification of the metabolites it contains. Certain
important nodes are depicted as triangles (non-hub connectors),
hexagons (connector hubs), and squares (provincial hubs). Interactions
between modules and nodes are depicted using lines, whith thickness
proportional to the number of actual links.
(Inset) Pajek-obtained representation of the entire metabolic network
of {\it E. coli} contains 473 metabolites and 574 links. Each node is
colored according to the ``main'' color of its module, as obtained
from the ``cartographic representation.''
}
\label{f-metab}
\end{figure}

First, we identify the functional modules in the different metabolic
networks (Fig.~\ref{f-metab}). Finding modules in metabolic networks
based on purely topological properties is an extremely important
task. For example, Schuster {\it et al.} have reported on the
impossibility of obtaining elementary flux modes \cite{schuster00}
from complete metabolic networks due to the combinatorial explosion of
the number of such modes \cite{schuster02}. Our algorithm identifies
an average of 15 different modules in each metabolic network---with a
maximum of 19 for {\it E. coli} and {\it H. sapiens}, and a minimum of
11 for {\it A. fulgidus}. As expected, the density of links within
each of the modules is significantly larger than between modules,
typically 100-1000 times larger (see Supplementary Information).

To assess how each of the modules is related to the pathways
traditionally defined in biology, we use the classification scheme
proposed in KEGG, which includes nine major pathways: carbohydrate
metabolism, energy metabolism, lipid metabolism, nucleotide
metabolism, amino acid metabolism, glycan biosynthesis and metabolism,
metabolism of cofactors and vitamins, biosynthesis of secondary
metabolites, and biodegradation of xenobiotics. Each metabolite in the
KEGG database is assigned to, at least, one pathway; thus, we can
determine to which pathways the metabolites in a given module
belong. We find that most modules contain metabolites mostly from one
major pathway. For example, in 17 of the 19 modules identified for
{\it E. coli}, more than one third of the metabolites belong to a
single pathway. Interestingly, some other modules---two in the case of
{\it E. coli}---cannot be trivially associated with a single
traditional pathway. These modules are typically central in the
metabolism and contain, mostly, metabolites that are classified in
KEGG as belonging to carbohydrate and amino acid metabolism.

Next, we identify the role of each metabolite. In
Fig.~\ref{f-roledef}b we show the roles identified in the metabolic
network of {\it E. coli}. Remarkably, other organisms display a
similar distribution of the nodes in the different roles, even though
they correspond to organisms that are very distant from an
evolutionary standpoint (see Supplementary Information). Role R1,
which contains ultra-peripheral metabolites with small degree and no
between-module links, comprises 76-86\% of all the metabolites in the
networks. This considerably simplifies the coarse-grained
representation of the network as these nodes do not need to be
identified separately. Note that this finding alone represents an
important step towards the goal of extracting scale-specific
information from complex networks.

\bigskip
\noindent
{\it Metabolite role and inter-species conservation.}
The information about modules and roles enables us to build a
``cartographic representation'' of the metabolic network of, for
example, {\it E. coli} (Fig.~\ref{f-metab}). This representation
enables us to recover relevant biological information. For instance,
we find that the metabolism is mostly organized around the module
containing pyruvate, which, in turn, is strongly connected to the
module whose hub is acetyl-CoA. These two molecules are key to connect
the metabolism of carbohydrates, amino acids, and lipids to the TCA
cycle from which ATP is obtained. These two modules are connected to
more peripheral ones by key metabolites such as D-glyceraldehyde
3-phosphate and D-fructose 6-phosphate (which connect to the glucose
and galactose metabolisms), D-ribose 5-phosphate (which connects to
the metabolism of certain nucleotides), and glycerone phosphate (which
connects to the metabolism of certain lipids).

Importantly, our analysis also uncovers nodes with key connector roles
that take part in only a small but fundamental set of reactions. For
example, N-carbamoyl-L-aspartate takes part in only three reactions
but is vital because it connects the pyrimidine metabolism, whose hub
is uracil, to the core of the metabolism through the alanine and
aspartate metabolism. The potential importance of such non-hub
connectors points to another consideration. It is a plausible {\it
hypothesis\/} that nodes with different roles are under different
evolutionary constraints and pressures. In particular, one expects
that nodes with structurally relevant roles are more necessary and
therefore more conserved across species.

To quantify the relation between roles and conservation, we define the
loss rate $p_{\rm lost}(R)$ (see Methods). Structurally relevant roles
are expected to have low values of $p_{\rm lost}(R)$ and vice
versa. Remarkably, we find that the different roles have, indeed,
different loss rates (Fig.~\ref{f-conservation}). As expected,
ultra-peripheral nodes (role R1) have the highest loss rate while
connector hubs (role R6) are the most conserved across all species
considered.

The results for the comparison of $p_{\rm lost}(R)$ for
ultra-peripheral nodes and connector hubs is illustrative, but hardly
surprising. The comparison of $p_{\rm lost}(R)$ for non-hub connectors
(role R3) and provincial hubs (role R5), however, yields a surprising
and remarkable finding. The metabolites in the provincial hubs class
have many within-module connections, sometimes as much as five
standard deviations more connections than the average node in the
module. Conversely, non-hub connector metabolites have few links
relative to other nodes in their modules---and fewer total connections
than the metabolites in role R5 (see Supplementary Figs.~S12b,c). The
links of non-hub connectors, however, are distributed among {\it
several different modules}, while the links of provincial hubs are
mainly within their modules. We find that non-hub connectors are
systematically and significantly more conserved than provincial hub
metabolites (Fig.~\ref{f-conservation}).

A possible explanation for the high degree of conservation of non-hub
connectors is the following. Connector nodes are responsible for
inter-module fluxes. These modules are, otherwise, poorly connected or
not connected at all to each other, so the elimination of connector
metabolites will likely have a large impact on the global structure of
fluxes in the network. On the contrary, the pathways in which
provincial hubs are involved may be backed up within the module, in
such a way that elimination of these metabolites may have a
comparatively smaller impact, which, in addition, would likely be
confined to the module containing the provincial hub.

Our results therefore point to the need to consider each complex
biological network as a whole, instead of focusing on local
properties. In protein networks, for example, it has been reported
that hubs are more essential than non-hubs
\cite{jeong01}. Notwithstanding the relevance of such a finding, our
results suggest that the global role of nodes in the network might be
a better indicator of their importance than degree~\cite{han04}.

Our ``cartography'' provides a scale-specific method to process the
information contained in the structure of complex networks, and to
extract knowledge about the function carried out by the network and
its constituents. An open question is how to adapt current
module-detection algorithms to networks with a hierarchical structure.

For metabolic networks, a comparatively well studied and well
understood case, our method allows us to recover firmly established
biological facts, and to uncover important new results, such as the
significant conservation of non-hub connector metabolites. Similar
results can be expected when our method is applied to other complex
networks that are not as well studied as metabolic networks. Among
those, protein interaction and gene regulation networks may be the
most significant.

%
%
\section*{Methods}

\subsection*{Modularity}

For a given partition of the nodes of a network into modules, the
modularity $M$ of this partition
is~\cite{newman03,newman04,guimera04c}
\begin{equation}
M\equiv\sum_{s=1}^{N_M}\left[\frac{l_{s}}{L}-
\left(\frac{d_s}{2L}\right)^2\right]\,,
\label{e-modularity}
\end{equation}
where $N_M$ is the number of modules, $L$ is the number of links in
the network, $l_{s}$ is the number of links between nodes in module
$s$, and $d_s$ is the sum of the degrees of the nodes in module
$s$. The rationale for this definition of modularity is the
following. A good partition of a network into modules must comprise
many within-module links and as few as possible between-module
links. However, if one just tries to minimize the number of
between-module links (or, equivalently, maximize the number of
within-module links) the optimal partition consists of a single module
and no between-module links. Equation (\ref{e-modularity}) addresses
this difficulty by imposing that $M=0$ if nodes are placed at random
into modules {\it or} if all nodes are in the same
cluster~\cite{newman03,newman04,guimera04c}.

The objective of a module identification algorithm is to find the
partition with largest modularity, and several methods have been
proposed to attain such a goal. Most of them rely on heuristic
procedures and use $M$---or a similar measure---only to assess their
performance. In contrast, we use simulated
annealing~\cite{kirkpatrick83} to find the partition with the largest
modularity.

\subsection*{Simulated annealing for module identification}

Simulated annealing\cite{kirkpatrick83} is a stochastic optimization
technique that enables one to find ``low cost'' configurations without
getting trapped in ``high-cost'' local minima. This is achieved by
introducing a {\it computational temperature} $T$. When $T$ is high,
the system can explore configurations of high cost while at low $T$
the system only explores low cost regions. By starting at high $T$ and
slowly decreasing $T$, the system descends gradually toward deep
minima, eventually overcoming small cost barriers.

When identifying modules, the objective is to maximize the modularity
and, thus, the cost is $C=-M$, where $M$ is the modularity as defined
in Eq.~(\ref{e-modularity}). At each temperature, we perform a number
of random updates and accept them with probability
\begin{equation}
p=\left\{ \begin{array}{lcl}
        1		&       \quad\mbox{if} &  C_f \le C_i\\
        \exp{\left(-\frac{C_f-C_i}{T}\right)} & \quad\mbox{if} &       C_f > C_i
        \end{array}\right. 
\end{equation}
where $C_f$ is the cost after the update and $C_i$ is the cost before
the update.

Specifically, at each $T$ we propose $n_i=fS^2$ individual node
movements from one module to another, where $S$ is the number of nodes
in the network. We also propose $n_c=fS$ collective movements, which
involve either the merging two modules or splitting a module. For $f$
we typically choose $f=1$. After the movements are evaluated at a
certain $T$, the system is cooled down to $T'=cT$, with $c=0.995$.

\subsection*{Within-module degree and participation coefficient}

Each module can be organized in very different ways, ranging from
totally centralized---with one or a few nodes connected to all the
others---to totally decentralized---with all nodes having similar
connectivities. Nodes with similar roles are expected to have similar
relative within-module connectivity. If $\kappa_{i}$ is the number of
links of node $i$ to other nodes in its module $s_i$,
$\overline{\kappa}_{s_i}$ is the average of $\kappa$ over all the
nodes in $s_i$, and $\sigma_{\kappa_{s_i}}$ is the standard deviation
of $\kappa$ in $s_i$, then
\begin{equation}
z_i = \frac{\kappa_i - \overline{\kappa}_{s_i}}{\sigma_{\kappa_{s_i}}}
\end{equation}
is the so-called $z$-score. The within-module degree $z$-score
measures how ``well-connected'' node $i$ is to other nodes in the
module.

Different roles can also arise because of the connections of a node to
modules other than its own. For example, two nodes with the same
$z$-score will play different roles if one of them is connected to
several nodes in other modules while the other is not. We define the
participation coefficient $P_i$ of node $i$ as
\begin{equation}
P_i=1-\sum_{s=1}^{N_M}\left(\frac{\kappa_{is}}{k_i} \right)^2
\end{equation}
where $\kappa_{is}$ is the number of links of node $i$ to nodes in
module $s$, and $k_i$ is the total degree of node $i$. The
participation coefficient of a node is therefore close to one if its
links are uniformly distributed among all the modules and zero if all
its links are within its own module.

\subsection*{Loss rate}

To quantify the relation between roles and conservation, we calculate
to which extent metabolites are conserved in the different species
depending on the role they play. Specifically, for a pair of species,
$A$ and $B$, we define the loss rate as the probability
$p(R_A=0|R_B=R) \equiv p_{\rm lost}(R)$ that a metabolite is not
present in one of the species ($R_A=0$) given that it plays role $R$
in the other species ($R_B=R$). Structurally relevant roles are
expected to have low values of $p_{\rm lost}(R)$ and vice versa.

%

\bigskip
\noindent
{\bf Acknowledgments}~~We thank L. Broadbelt, V. Hatzimanikatis,
A.~A. Moreira, E. T. Papoutsakis, M. Sales-Pardo, and D.~B. Stouffer
for stimulating discussions and helpful suggestions, and H. Ma and
A. P. Zeng for providing us with their metabolic networks'
database. R.G. thanks the Fulbright Program and the Spanish Ministry
of Education, Culture \& Sports. L.A.N.A. gratefully acknowledges the
support of a Searle Leadership Fund Award and of a NIH/NIGMS K-25
award.

\clearpage

\end{document}